\documentclass[manuscript]{aastex}

\shorttitle{Horizontal Divergent Flow in the Sun}
\shortauthors{Toriumi et al.}

\begin{document}


\title{Statistical Analysis of the Horizontal Divergent Flow
  in Emerging Solar Active Regions}


\author{Shin Toriumi$^{1}$, Keiji Hayashi$^{2,3}$, and Takaaki Yokoyama$^{4}$}
\affil{$^{1}$National Astronomical Observatory of Japan, 2-21-1 Osawa, Mitaka, Tokyo 181-8588, Japan}
\email{shin.toriumi@nao.ac.jp}
\affil{$^{2}$W. W. Hansen Experimental Physics Laboratory,
  Stanford University,
  Stanford, CA 94305, USA}
\affil{$^{3}$Key Laboratory of Solar Activity,
  National Astronomical Observatories of China,
  Chinese Academy of Sciences, Beijing, 100012, China}
\affil{$^{4}$Department of Earth and Planetary Science,
  University of Tokyo,
  7-3-1 Hongo, Bunkyo-ku, Tokyo
  113-0033, Japan}




\begin{abstract}
Solar active regions (ARs)
are thought to be formed
by magnetic fields
from the convection zone.
Our flux emergence simulations
revealed that
a strong horizontal divergent flow (HDF)
of unmagnetized plasma
appears at the photosphere
before the flux begins to emerge.
In our earlier study,
we analyzed HMI data
for a single AR
and confirmed presence
of this precursor plasma flow
in the actual Sun.
In this paper,
as an extension of our earlier study,
we conducted a statistical analysis
of the HDFs
to further investigate their characteristics
and better determine the properties.
From {\it SDO}/HMI data,
we picked up 23 flux emergence events
over a period of 14 months,
the total flux of which ranges
from $10^{20}$ to $10^{22}\ {\rm Mx}$.
Out of 23 selected events,
6 clear HDFs were detected
by the method
we developed in our earlier study,
and 7 HDFs detected
by visual inspection
were added to this statistic analysis.
We found that
the duration of the HDF
is on average 61 minutes
and the maximum HDF speed
is on average $3.1\ {\rm km\ s}^{-1}$.
We also estimated
the rising speed of the subsurface magnetic flux
to be $0.6$--$1.4\ {\rm km\ s}^{-1}$.
These values are highly consistent
with our previous one-event analysis
as well as our simulation results.
The observation results lead us
to the conclusion that
the HDF is rather a common feature
in the earliest phase of AR emergence.
Moreover,
our HDF analysis has capability
of determining the subsurface properties
of emerging fields
that cannot be directly measured.
\end{abstract}





\section{Introduction
  \label{sec:introduction}}

Active regions (ARs)
including sunspots are
one of the most prominent features
in the Sun.
These regions are highly magnetized and,
through magnetic reconnection and instability,
they may produce catastrophic eruptions
known as flares and CMEs.
\citet{par55} suggested that
ARs are created
by rising magnetic fields
from the deep convection zone
(flux emergence).

In the last decades,
numerical simulations
applying
a number of different
approximations
have widely been carried out
and they successfully explained
many observational characteristics
found in ARs.
For instance,
simulations using the thin-flux-tube approximation
revealed that
Coriolis force
acting on the rising field
is responsible for
the production of
AR tilts and asymmetries
\citep{dsi93,fan93}.
Emerging flux
in a turbulent convective envelope
was addressed
employing the anelastic approximation
\citep[e.g.,][]{jou09}.
However,
these assumptions become inappropriate
in the upper convection zone
around $-20\ {\rm Mm}$
due to the drastic changes
in physical parameters
\citep{fan09}.

On the other hand,
emergence simulations 
from the surface layer
have elucidated
the dynamical features
of newly emerging flux regions.
Two-dimensional magnetohydrodynamic (2D MHD) simulations
by \citet{shi89} reproduced
the rising motion of
arch filament system \citep[AFS:][]{bru69}
and the supersonic downflows
along the magnetic fields.
\citet{mag01} calculated
the 2D cross-sectional evolutions
of twisted flux tubes,
while \citet{fan01} compared
her 3D simulation results
with AR observation
by \citet{str96}.
3D MHD simulations
of rising, twisted flux tubes
by \citet{abb00} show
a large, diverging horizontal flow field
near the apex of the tube
as it approaches the upper boundary
(see, e.g., Figure 14 of that paper).

Recently,
in \citet{tor12a},
we conducted
a 3D MHD simulation
of a rising magnetic flux tube
in a larger scale
from a depth of $-20\ {\rm Mm}$
in the convection zone,
where the approximations break down,
to the solar corona
through the photosphere.
We found that
the flux tube
decelerates temporarily
beneath the surface of the Sun
before it restarts emergence
into the upper atmosphere
(``two-step emergence'' model).
In this simulation,
the flux tube goes up
the solar interior
at a rate of $\sim 1\ {\rm km\ s}^{-1}$
and,
due to the external
density and pressure stratification,
it begins to expand.
Moreover,
because of the expansion,
the tube starts to push up
unmagnetized plasma
on its apex.
The plasma becomes
trapped between the rising tube
and the isothermally-stratified
(i.e., convectively-stable) photosphere
and, finally,
it escapes horizontally
around the surface layer
just before the start
of the flux emergence
in the photosphere,
which we call
the horizontal divergent flow (HDF).

In \citet{tor13a},
we extended the above simulation
and carried out
a parametric survey,
finding that
the driving force of the HDF
is the horizontal pressure gradient.
We also found that
the duration of the HDF
(the time gap between
the start of the HDF
and that of the flux emergence)
ranges 30--45 minutes,
while the horizontal velocity
is typically several km s$^{-1}$,
a fraction of
the photospheric sound speed
($\sim 10\ {\rm km\ s}^{-1}$).
In the convective emergence simulations
that solve radiative MHD,
similar outflows are also observed
in the earliest moments
of flux emergence
\citep{che10}.
\citet{rem14} reported that
the flow speed reaches up to
$2\ {\rm km\ s}^{-1}$.

The HDF is detected
in the actual Sun.
In \citet{tor12b},
we investigated
NOAA AR 11081,
which emerged
closer to the northwestern limb
on 2010 June 11,
and found that
the HDF precedes
the appearance of magnetic flux.
The HDF continued for 103 minutes,
while its speed
was $0.6$--$1.5\ {\rm km\ s}^{-1}$,
up to $2.3\ {\rm km\ s}^{-1}$.
These values are comparable
to the simulation results,
which supports
the ``two-step emergence'' model
based on the numerical experiments.

Although the previous detection
is a promising result,
we observed the HDF
only in a single emergence event.
Therefore,
we need to repeat our analysis
in many more events
to further support
the theoretical model.
Also,
the parameters of the HDF,
such as the duration
and the maximum velocity,
may depend on
the properties
of the subsurface magnetic field
that pushes up the plasma.
By conducting numerical simulations
of flux tube emergence,
we found that
these parameters
actually depend on the field strength
and the twist intensity
of the initial flux tube
\citep{tor13a}.
Therefore,
by analyzing HDFs
in a larger ensemble
of emergence events,
we may be able to investigate
the physical aspects
of the subsurface flux
such as the rising speed.

In this paper,
we report
on the statistical analysis
of the HDFs
in 23 flux emergence events.
The purpose of this study is
to detect HDFs
in a larger ensemble
of AR data
and investigate their characteristics.
Moreover,
based on the HDF analysis,
we further aim
to derive any physical properties
of rising magnetic fields
in the solar interior.
The rest of the paper
is organized
as follows.
In the next section,
we provide the data selection and reduction.
In Section \ref{sec:analysis},
we analyze the HDFs
and show the results.
Section \ref{sec:discussion}
is dedicated to
discussing the results
and deriving
the physical properties
of rising flux tubes
in the convection zone.
Finally,
in Section \ref{sec:summary},
we summarize this paper.

\section{Data Selection and Reduction
  \label{sec:data}}

\subsection{Data Selection}

In this study,
we used observational data
taken by the Helioseismic and Magnetic Imager
\citep[HMI:][]{sche12,scho12}
on board the {\it Solar Dynamics Observatory}
({\it SDO}).
Thanks to its continual,
full-disk observation,
we are able to follow
the AR evolution
from its earliest stage.
We searched for
flux emergence events
that occurred during the period
from May 2010 to June 2011 (14 months),
which is the solar minimum
between Solar Cycles 23 and 24
after the launch
of the {\it SDO} spacecraft
in February 2010.
The reason of choosing this period
is to obtain ``clear'' events,
that is,
a flux emergence
into quiet Sun
with less preexisting magnetic fields.
As the Sun steps into the active phase,
it may become difficult
to distinguish magnetic flux
of the newly emerged fields
from that of the preexisting fields,
which may be the remnant
of previously emerged fields.
During this period,
we selected emergence events
in the area with a heliocentric angle
$\theta\leq 60^{\circ}$
to keep the quality of HMI data.
Also,
we picked up
only ARs that emerged
in the eastern hemisphere
of the solar disk
in order to monitor
the AR evolution
as long as possible
from their births.
Through these criteria,
we obtained
23 emergence events
in 21 ARs
including 2 ephemeral ARs
\citep{har73}.

\subsection{Data Reduction
  \label{sec:reduction}}

For each target AR,
we made tracked data cubes
of the Doppler velocity
and the line-of-sight (LoS) magnetogram,
both having a pixel size of
0.03 heliographic degree
($\simeq 0.5\ {\rm arcsec} \simeq 360\ {\rm km}$)
with a $512\times 512$ pixel field-of-view (FoV),
and a 12 minute cadence
with a 7 day duration.
Also,
for each emergence event,
we made tracked data cubes
of Dopplergram and LoS magnetogram
with the same resolution
and FoV
but with a cadence of $45\ {\rm s}$
and a duration of 36 hr.
For each data cube,
we applied Postel projection.
%
%

In order to eliminate
the effects of the rotation of the Sun
and the orbital motion of the {\it SDO} spacecraft,
and to reduce the east-west trend
(due to the spherical geometry
of the Sun)
from the Dopplergram,
we constructed additional background data
and subtracted the background
from each Doppler frame.
The reduction procedure is
as follows
(based on the method
by \citealt{gri07}).
\begin{enumerate}
\item To estimate the background level
  at the top of the FoV,
  we first averaged
  the topmost ten-pixel rows
  to obtain a strip of data
  running in the east-west direction.
  The strip was then linearly smoothed.
  We repeated the same for the bottom,
  to have background data
  with only the top and the bottom rows
  filled in.
\item We performed
  a linear interpolation
  between the upper and lower pixels
  of the columns
  to produce full background data.
\item By subtracting
  this background data
  from each frame,
  we obtained trend-free Doppler data.
\end{enumerate}
In addition,
a 10-minute running average
was applied to the 36 hr Dopplergrams and magnetograms
to smooth out
rapid fluctuations
that may not be related
to the HDF
(5-minute oscillation,
surface gravity wave, etc.).

\section{Analysis
  \label{sec:analysis}}

\subsection{Properties of ARs
  \label{sec:analysis:properties}}

In this subsection,
we analyze the properties
of 21 ARs.
For each 7-day magnetogram
of the target AR,
we measured
the maximum total unsigned flux,
the maximum unsigned flux growth rate,
and the maximum footpoint separation.
In the magnetogram,
we first set a box surrounding
the emerging region,
and measured the total unsigned flux
$\Phi=\int_{S}|B|\, dS$
inside the box
and its time derivative $d\Phi/dt$,
where $B$ is
the LoS magnetic field strength
of each pixel,
$S$ is the box size,
$dS$ is the area of a pixel element,
and $t$ is the time.
We applied 120 and 240 minute smoothings
for $\Phi$ and $d\Phi/dt$,
respectively,
and recorded their maximum values,
$\max{(\Phi)}$ and $\max{(d\Phi/dt)}$.
Also,
within the box in each frame,
we measured
the flux-weighted center
of each polarity
\begin{eqnarray}
  (x_{\pm}, y_{\pm})
  = \left(
     \frac{\sum xB_{\pm}}{\sum B_{\pm}},
     \frac{\sum yB_{\pm}}{\sum B_{\pm}}
    \right),
\end{eqnarray}
where $B_{\pm}$ is the field strength
of each pixel and
$+$ ($-$)
is for positive (negative) polarity,
and evaluated the footpoint separation
between both centers,
\begin{eqnarray}
  d_{\rm foot}
  = \sqrt{ (x_{-}-x_{+})^{2} + (y_{-}-y_{+})^{2} }.
\end{eqnarray}
Here,
for the evaluation
of the flux-weighted centers,
we only used the pixels
with $|B|\geq 200\ {\rm G}$
to keep the data quality.
Then,
we recorded the maximum separation,
$\max{(d_{\rm foot})}$.

\begin{figure}
  \begin{center}
    \includegraphics[width=70mm]{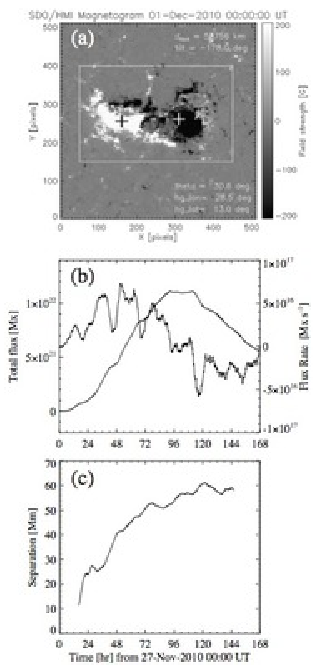}
  \end{center}
  \caption{
    An example of AR data
    from the list
    in Appendix \ref{sec:catalog}.
    (a) HMI magnetogram
    of NOAA AR 11130
    (event \#7).
    Black and white crosses are
    the flux-weighted centers
    of positive and negative polarities,
    respectively,
    which are calculated
    inside the rectangular box.
    (b) Temporal evolutions of
    the total unsigned flux
    ($\Phi$: thick)
    and its time derivative
    ($d\phi/dt$: thin).
    (c) Temporal evolution
    of the footpoint separation
    ($d_{\rm foot}$).}
  \label{fig:properties}
\end{figure}

Figure \ref{fig:properties}
is an example
of AR data.
In panel (a),
the HMI magnetogram of NOAA AR 11130
(event \#7)
is shown
and the flux-weighted center
of each polarity
is indicated with a cross sign.
Panel (b) is for
the temporal evolutions
of the total unsigned flux
($\Phi$)
and
the flux growth rate
($d\Phi/dt$).
One interesting characteristic
of this figure
is the several peaks
in the
flux rate curve,
which may indicate that
the emerging flux is bifurcated
and consists of separate flux bundles
\citep{zwa85}.
Panel (c) shows
the footpoint separation
($d_{\rm foot}$).
The maximum total flux,
the maximum flux rate,
and the maximum separation
of this AR
are $1.1\times 10^{22}\ {\rm Mx}$,
$7.3\times 10^{16}\ {\rm Mx\ s}^{-1}$,
and $61.1\ {\rm Mm}$,
respectively.
The corresponding figures
of all the analyzed ARs
are listed
in Appendix \ref{sec:catalog}.

The obtained properties
of the target regions
are summarized
in Table \ref{tab:properties}.
As can be seen in this table,
the total flux ranges
from $5.6\times 10^{20}$
to $2.3\times 10^{22}\ {\rm Mx}$,
while
the flux growth rate ranges
from $1.3\times 10^{16}$
to $1.7\times 10^{17}\ {\rm Mx\ s}^{-1}$
and the footpoint separation
ranges from $20.5$
to $89.3\ {\rm Mm}$.
Note that,
in order to keep the data quality,
the physical values
of each AR
in this table
are measured
under the condition that
the heliocentric angle is
$\theta \leq 60^{\circ}$.
We should also take care of the fact that,
because of the diffusion of the sunspots,
footpoint separations of some ARs
may become larger
even after the ARs
complete their growths.
Regarding quadrupolar regions
such as NOAA AR 11158 (event \#16),
we measure the separation
not between the most distant footpoints
but between the two representative flux-weighted centers
simply calculated
from the entire AR.

\subsection{Detection of the HDF
  \label{sec:analysis:hdf}}


First,
for each 36 hr data
of the 23 emergence events,
we measured
the start of the HDF
in the Dopplergram
(HDF start: $t_{\rm HDF}$)
and the start of the flux emergence
in the LoS magnetogram
(emergence start: $t_{\rm FE}$).
%
Here,
we used the method developed
in \citet{tor12b}:
We plotted the histograms
of the Doppler velocity
and the absolute LoS magnetic field strength
inside the square
that surrounds the emergent area.
By focusing on the residuals of the histograms
from their reference quiet-Sun profiles,
i.e., the histograms before the emergence,
we investigated the temporal evolutions
of the Doppler and magnetic fields.
The HDF start (emergence start)
was determined
as the time when
the high-speed component
of the Doppler residual
(strongly-magnetized component
of the magnetic residual)
exceeded
the one standard deviation ($1\sigma$) level
of the reference profile.
For the details of the method,
see Section 3.2 of \citet{tor12b}.
In the present study,
we used the square of the size
of $64\times 64$ pixels
($\sim 23\ {\rm Mm}\times 23\ {\rm Mm}$).
The ranges for the high-speed and strongly-magnetized components
depend on the emergence event,
typically being
[$-1.5\ {\rm km\ s}^{-1}$, $-1\ {\rm km\ s}^{-1}$]
and [$1\ {\rm km\ s}^{-1}$, $1.5\ {\rm km\ s}^{-1}$]
for the Dopplergram
and [$200\ {\rm G}$, $300\ {\rm G}$]
for the magnetogram.

\begin{figure}
  \begin{center}
    \includegraphics[width=145mm]{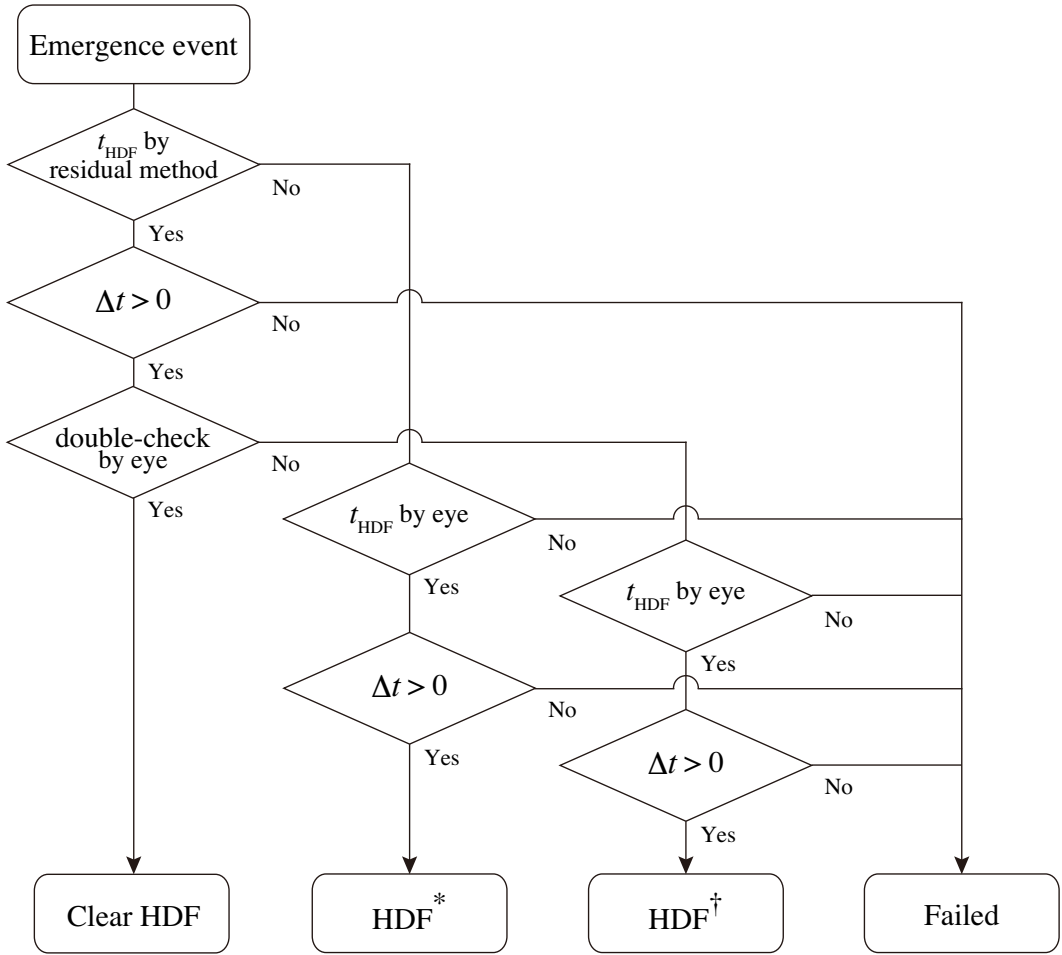}
  \end{center}
  \caption{Flowchart of the HDF detection.
    Here,
    $t_{\rm HDF}$ and $t_{\rm FE}$ denote
    the HDF start and the emergence start,
    respectively,
    while $\Delta t=t_{\rm FE}-t_{\rm HDF}$.
    In every event,
    $t_{\rm FE}$ is determined
    and thus this process
    is not shown
    in the chart.}
  \label{fig:flow}
\end{figure}

The analysis procedure
of the HDF detection
is shown as a flowchart
in Figure \ref{fig:flow}.
Since
in every emergence event
we were able to
determine $t_{\rm FE}$
using the above residual method,
we skipped this process
in the chart.
In this procedure,
if $t_{\rm HDF}$ was once defined
by the residual method,
we then calculated
the time gap
between these two times
(HDF duration: $\Delta t=t_{\rm FE}-t_{\rm HDF}$).
If $\Delta t>0$,
we double-checked the Doppler images
by visual inspection
and determined
if the HDF was clear
(clear HDF) or not.
If $\Delta t\leq 0$,
we defined that
the HDF detection was failed.
If $t_{\rm HDF}$ was not defined
by the residual method,
we determined $t_{\rm HDF}$
by visual inspection
and, again,
if $\Delta t>0$,
we defined that
the HDF was detected
(HDF by eye:
indicated with asterisk).
If the result of the double-check
was negative,
we also determined $t_{\rm HDF}$ by eye
and checked if $\Delta t>0$
(HDF by eye:
indicated with dagger).

The results of the detection
are shown
in Table \ref{tab:hdf}.
In 6 emergence events
out of the entire 23 events,
we observed clear HDFs,
that is, $t_{\rm HDF}$ was detected
by the residual method,
satisfied $\Delta t>0$,
and passed the double-check process.
The temporal evolutions
of the Dopplergram and magnetogram
for the 6 clear events,
along with the corresponding continuum images,
are shown
in Figure \ref{fig:clear1}
in Appendix \ref{sec:clear}.
In another 7 events,
we detected HDFs
by visual inspection
instead of the residual method.
Thus, in total,
HDFs were found
in 13 events, or,
56.5\% of all the analyzed events.

In the 6 clear events,
the HDF duration $\Delta t$
ranges from 5 to 106 minutes
(the average $61.0$ minutes
and the median $56.0$ minutes).
For these events,
we also evaluated
the maximum horizontal speed
from the Doppler velocity,
$V_{\rm D}$.
During $\Delta t$,
we applied a slit
with a thickness of 5 pixels
to the Dopplergrams
and averaged over 5 pixels,
and measured the largest absolute Doppler velocity.
Here, the slit
is parallel to the separation
of both magnetic polarities,
centered at the middle of the emergent region.
Using the heliocentric angle $\theta$,
the maximum horizontal velocity
can be obtained by
\begin{eqnarray}
  \max{(|V_{\rm h}|)}
  = \frac{\max{(|V_{\rm D}|)}}{\sin{\theta}}.
\end{eqnarray}
Table \ref{tab:clear}
shows the results
of this analysis.
The maximum HDF speed
ranges from $1.8$ to $4.4\ {\rm km\ s}^{-1}$
(the average
$3.1\ {\rm km\ s}^{-1}$
and the median $3.4\ {\rm km\ s}^{-1}$).
The obtained durations
and the horizontal speeds
will be discussed
in Section \ref{sec:discussion:physical}.

\section{Discussion
  \label{sec:discussion}}

\subsection{HDF Detection}

\begin{figure}
  \begin{center}
    \includegraphics[width=150mm]{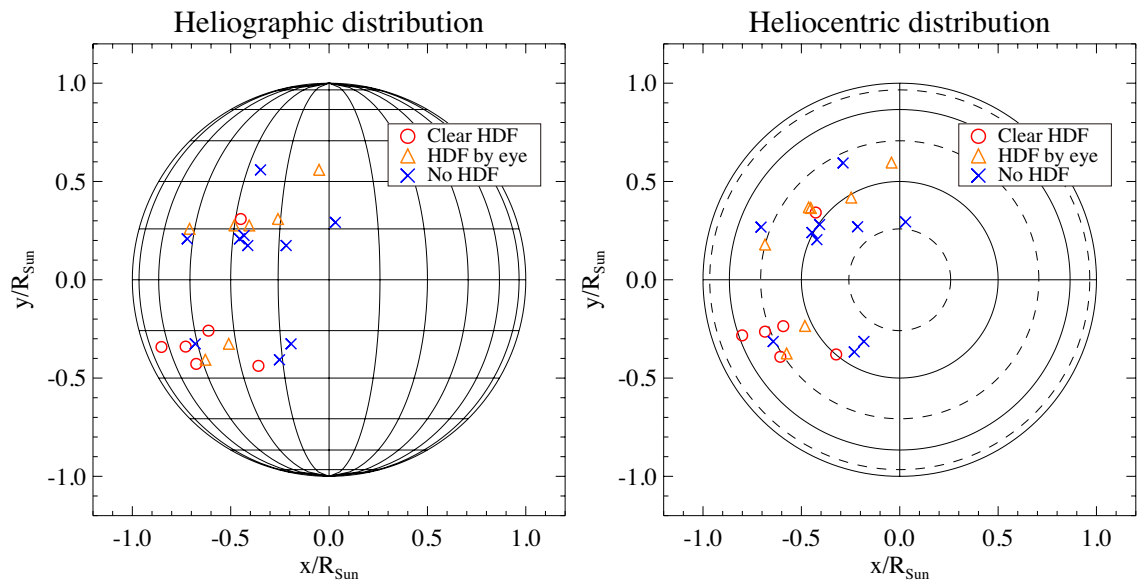}
  \end{center}
  \caption{
    Distribution of the flux emergence events.
    (Left) Heliographic coordinates,
    where the Sun's rotation axis is fixed
    onto the $x$--$y$ plane.
    (Right) Heliocentric coordinates,
    where the LoS is perpendicular
    to the $x$--$y$ plane,
    i.e., $z$-axis is toward the Earth.}
  \label{fig:distribution}
\end{figure}

In the previous section,
we analyzed 23 flux emergence events
and detected 6 clear HDFs.
Also, we found 7 more HDFs
by visual inspection.
Figure \ref{fig:distribution}
shows the distribution
of the events
in heliographic and heliocentric coordinates.
From the heliographic map,
one may find
that newly emerging ARs
are distributed
in the mid-latitude bands
of both hemispheres
(latitude ranging approximately
from $\pm 15^{\circ}$
to $\pm 30^{\circ}$).
Also,
from the heliocentric map,
one can see that
all clear HDFs are detected
in the range of $\theta>30^{\circ}$.

In the remaining 10 events,
we did not find HDFs.
The possible reasons
for the failed detections
are as follows.
\begin{itemize}
\item If the flux emergence occurs
at the location
too close to the disk center,
the HDF may not appear
in the Dopplergrams
because of the projection.
In fact,
all 6 clear events
are located
away from the disk center
($\theta > 30^{\circ}$),
while 7 out of 10 failed events
are closer to the center
($\theta\leq 30^{\circ}$).
\item The HDF may be stronger
in the direction
of the separation
of the positive and negative polarities
\citep[see, e.g., Figure 4 of][]{tor12a}.
When the footpoint separation
on the solar disk
is perpendicular
to the axis
from the disk center
to the target AR,
the HDF may not be seen,
since it has less Doppler component.
This effect seems to be responsible
for the failed detection
in event \#11.
\item The smaller ARs
may not have coherency or energy
enough to push up
the sufficient amount of plasma
that can be observed as HDF.
In the two ephemeral ARs
(events \#8 and \#21),
we did not observe HDFs
in both cases.
\item If the flux emerges
into a preexisting field,
it is difficult to separate the HDF
from the Doppler component
caused by the preexisting field
(event \#23).
\end{itemize}

\subsection{Comparison
  with Numerical Models
  \label{sec:discussion:physical}}

For the 6 clear HDF events,
we observed that
the HDF duration
is on average 61 minutes.
This value is comparable to
the duration of 103 minutes
previously measured
in \citet{tor12b}.
This value is also consistent with
30--45 minutes
obtained from a parameter survey
of the flux emergence simulations
in \citet{tor13a}.
According to the simulations,
the HDF duration
is comparable to
the elapse time
from the deceleration
of the rising flux
at the top convection zone
to the start of further emergence
into the upper atmosphere.
In other words,
the time gap of 61 minutes
indicates the waiting time
for the secondary emergence
in the ``two-step emergence'' model
(Section \ref{sec:introduction}).
Note that,
in the actual Sun,
thermal convection
is continuously excited
around the surface layer
and thus the situation
may be more complex.

The maximum HDF speed
of the 6 clear events
is on average $3.1\ {\rm km\ s}^{-1}$,
which is again comparable
to $2.3\ {\rm km\ s}^{-1}$
obtained in the previous observation
\citep{tor12b}.
According to our simulation
\citep{tor13a},
the HDF is driven
by the pressure gradient
and the maximum velocity
is several km s$^{-1}$,
which agrees with
the present observation results.
In the convective emergence simulation,
such a horizontally-diverging flow is also seen
in the earliest phase
of the flux emergence
\citep{che10}.
According to the simulation
by \citet{rem14},
the flow speed is up to
$2\ {\rm km\ s}^{-1}$,
which is comparable
to our observation of
$3.1\ {\rm km\ s}^{-1}$.

\subsection{Investigation of the Subsurface Magnetic Fields
  \label{sec:discussion:subsurface}}

In this section,
we investigate
the subsurface rising magnetic fields
that push up the unmagnetized plasma,
which is observed as an HDF.

\begin{figure}
  \begin{center}
    \includegraphics[width=80mm]{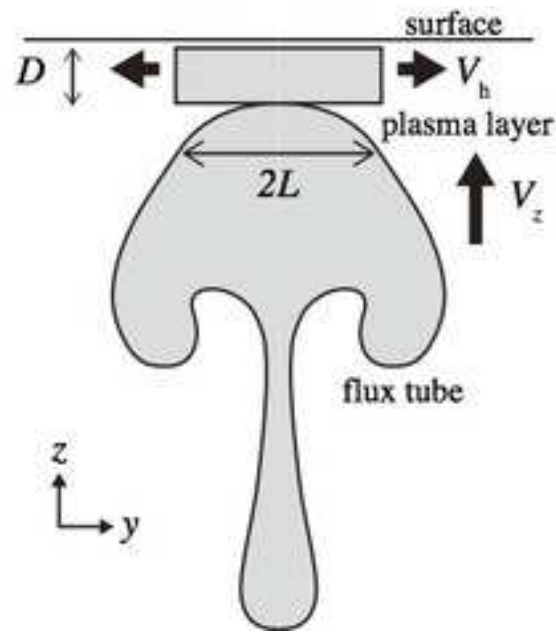}
  \end{center}
  \caption{2D cross-sectional model
    of a rising flux tube
    and a plasma layer
    ahead of the tube.
    Figure reproduced
    from \citet{tor13a}.}
  \label{fig:model}
\end{figure}

First,
we consider a simple 2D model
of the emerging magnetic flux
as illustrated
in Figure \ref{fig:model}.
Here,
we assume that
the rising speed of the magnetic flux $V_{z}$
is of the order of Alfv\'{e}n speed $V_{\rm A}$
\citep{par75},
namely,
\begin{eqnarray}
  V_{z} = \alpha V_{\rm A}
  =\alpha \frac{B}{\sqrt{4\pi\rho}},
  \label{eq:vz}
\end{eqnarray}
where $B$ and $\rho$
are the field strength
and the plasma density.
For simplicity,
we here assume
the factor $\alpha$ to be unity.
From the mass conservation
of the HDF,
we obtain
\begin{eqnarray}
  V_{\rm h} = \frac{L}{D}V_{z}
  = \frac{L}{D} \frac{B}{\sqrt{4\pi\rho}},
  \label{eq:masscons}
\end{eqnarray}
where $V_{\rm h}$, $L$, and $D$,
are the horizontal speed,
the lateral extension,
and the thickness
of the HDF,
respectively.
Also, the flux growth rate
when the flux appears at the surface
can be written as:
\begin{eqnarray}
  \frac{d\Phi}{dt}
  = 2LV_{z}B
  = 2L \frac{B^{2}}{\sqrt{4\pi\rho}},
  \label{eq:fluxrate}
\end{eqnarray}
where $\Phi$ is the magnetic flux.
Combining Equations (\ref{eq:masscons})
and (\ref{eq:fluxrate}),
we get
\begin{eqnarray}
  V_{\rm h} = \frac{(L/2)^{1/2}}{D(4\pi\rho)^{1/4}}
  \left[
    \frac{d\Phi}{dt}
  \right]^{1/2}.
  \label{eq:relation}
\end{eqnarray}
In this equation,
the horizontal speed $V_{\rm h}$
and the flux growth rate $d\Phi/dt$
can be measured
from the observational data,
which are summarized
in Table \ref{tab:clear}.

\begin{figure}
  \begin{center}
    \includegraphics[width=150mm]{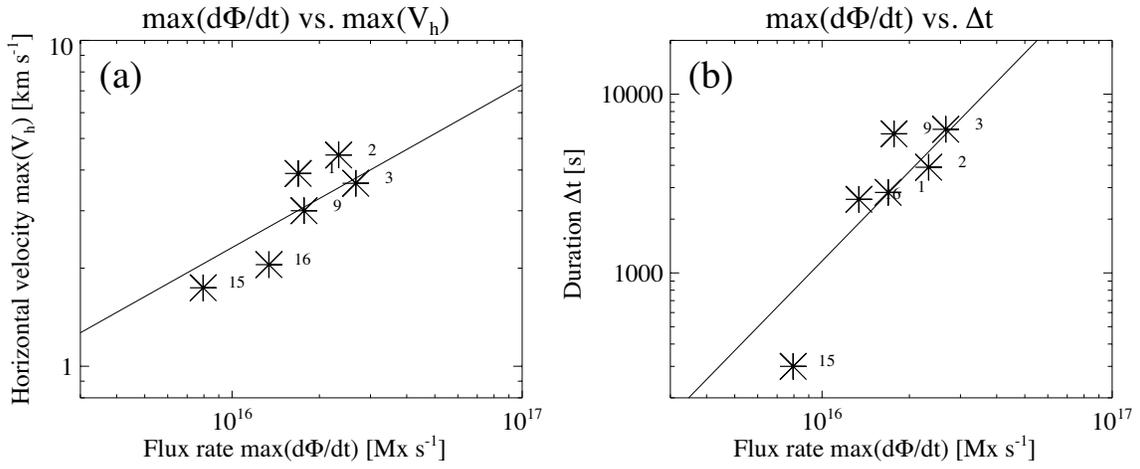}
  \end{center}
  \caption{
    (a) Maximum flux growth rate,
    $\max{(d\Phi/dt)}$,
    and maximum horizontal speed,
    $\max{(V_{\rm h})}$,
    for the 6 clear HDF events.
    The solid line is
    the fitted function
    $\max{(V_{\rm h})}=C_{1}\times
    [\max{(d\Phi/dt)}]^{1/2}$.
    (b) Maximum flux growth rate,
    $\max{(d\Phi/dt)}$,
    and HDF duration,
    $\Delta t$,
    for the 6 clear HDF events.
    The solid line is
    the fitted function
    $\Delta t= C_{2}\times[\max(d\Phi/dt)]^{C_{3}}$.
    In both panels,
    the number right to each asterisk
    represents the event number.
  }
  \label{fig:fitting}
\end{figure}

In Figure \ref{fig:fitting}(a),
we plot $\max{(d\Phi/dt)}$ and $\max{(V_{\rm h})}$,
measured during $\Delta t$,
for the 6 clear HDF events.
Here,
we fit a function
$\max{(V_{\rm h})}=C_{1}\times
[\max{(d\Phi/dt)}]^{1/2}$
in the diagram.
The obtained constant $C_{1}$
is comparable to the coefficient
in Equation (\ref{eq:relation}).
The result of the fitting is
\begin{eqnarray}
  C_{1}
  =\frac{(L/2)^{1/2}}{D(4\pi\rho)^{1/4}}
  = 2.3\times 10^{-3}.
  \label{eq:fitresult}
\end{eqnarray}
In the fitting process,
$V_{\rm h}$ is measured
in the unit of ${\rm cm\ s}^{-1}$.
From Equation (\ref{eq:fitresult}),
by assuming
$L=5\ {\rm Mm}$
and $\rho=2.5\times 10^{-7}\ {\rm g\ cm}^{-3}$,
we obtain the thickness of the HDF,
$D=1.6\ {\rm Mm}$.

Finally,
by inserting $D=1.6\ {\rm Mm}$
into Equation (\ref{eq:masscons}),
the rising speed of the magnetic flux,
$V_{z}$, is evaluated
for each HDF event,
which is summarized
in the rightmost column
of Table \ref{tab:clear}.
The rising speed
of the magnetic flux ranges
from $0.6$ to $1.4\ {\rm km\ s}^{-1}$.
This value is comparable to
our simulation results
of $\sim 1\ {\rm km\ s}^{-1}$
in \citet{tor13a}
and other simulations
of the flux emergence
within the convection zone
\citep[see][]{fan09}.
This obtained speed
is also comparable to
the helioseismic detections
of rising magnetic fields
\citep[e.g.,][]{ilo11,tor13b}.

Figure \ref{fig:fitting}(b)
shows the relation between
the maximum flux growth rate,
$\max{(d\Phi/dt)}$,
and the HDF duration,
$\Delta t$.
According to Equation (\ref{eq:fluxrate}),
$d\Phi/dt$ is proportional
to the square of $B$.
The linear trend
in Figure \ref{fig:fitting}(b)
is the fitted function
$\Delta t= C_{2}\times[\max(d\Phi/dt)]^{C_{3}}$
and the best fitting parameters are
$C_{2}=2.7\times 10^{-24}$
and $C_{3}=1.7$.
Thus,
the observation indicates
$\Delta t\propto B^{3.4}$.
In the parameter survey
of the flux emergence simulations
\citep[see Figure 4 of][]{tor13a},
we found that,
when the initial tube is stronger
(field strength $\gtrsim 30\ {\rm kG}$),
the HDF duration $\Delta t$
is inversely proportional
to the field strength
of the initial flux tube
at a depth of $-20\ {\rm Mm}$.
On the other hand,
when the tube is weaker
(field strength $\lesssim 30\ {\rm kG}$),
it deviates from the inverse variation
and has a positive correlation
with the field strength.
Therefore,
the observed positive correlation
of $\Delta t\propto B^{3.4}$
hints that
the field strength
of the rising flux
in the deeper convection zone
($\sim -20\ {\rm Mm}$)
has a field strength
of $\lesssim 30\ {\rm kG}$.

\subsection{Correlation with AR Parameters
  \label{sec:discussion:other}}

\begin{figure}
  \begin{center}
    \includegraphics[width=130mm]{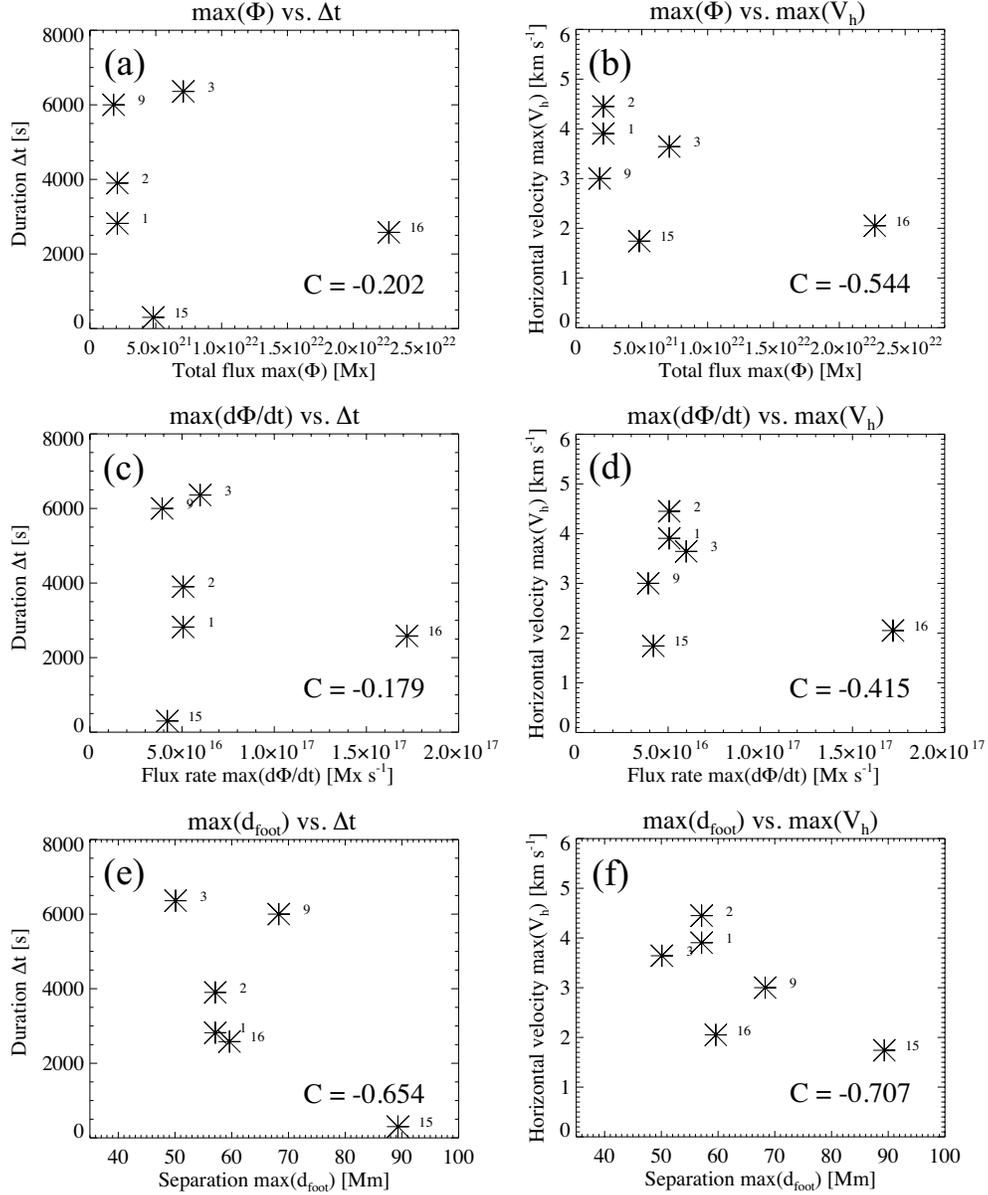}
  \end{center}
  \caption{
    Correlations
    between HDF parameters
    (HDF duration, $\Delta t$,
    and maximum HDF speed,
    $\max{(V_{\rm h})}$)
    and AR parameters
    (maximum total unsigned flux, $\max{(\Phi)}$,
    maximum flux rate, $\max{(d\Phi/dt)}$,
    and maximum footpoint separation, $\max{(d_{\rm foot})}$).
    Correlation coefficient $C$
    is indicated
    in the bottom right
    of each panel.
    Note that these parameters
    are measured
    from the 7-day magnetograms,
    under the condition
    of $\theta\leq 60^{\circ}$.
  }
  \label{fig:arparam}
\end{figure}

Figure \ref{fig:arparam}
summarizes the correlations
between HDF parameters
(HDF duration, $\Delta t$,
and maximum HDF speed, $\max{(V_{\rm h})}$)
and AR parameters
(maximum total unsigned flux, $\max{(\Phi)}$,
maximum flux rate, $\max{(d\Phi/dt)}$,
and maximum footpoint separation, $\max{(d_{\rm foot})}$)
for 6 clear HDF events.
Note that the AR parameters
including $\max{(d\Phi/dt)}$
were measured
not during the HDF duration
(i.e., $\Delta t$)
but during the entire AR evolution
using 7-day magnetogram.
The measurement is limited
by the condition
that the heliocentric angle,
$\theta$,
is $\leq 60^{\circ}$
(see Section \ref{sec:analysis:properties}).

%
In this figure,
the best correlation
is $-0.7$ of panel (f),
the correlation between
the maximum footpoint separation,
and the maximum HDF speed.
According to the simulation results
of \citet{tor13a},
the maximum HDF speed
is proportional
to the initial field strength
(see Figure 4(c) of that paper).
On the other hand,
when the field is stronger,
the total flux is expected to be larger,
and thus the AR size,
or, the footpoint separation
may also become larger.
Therefore,
the maximum separation
and the maximum horizontal speed
are expected to have
a positive correlation.
However,
the observational result
of Figure \ref{fig:arparam}(f)
is totally opposite from this expectation.
The absolute correlation coefficients
of other panels
are at most,
or less than $0.6$.
The AR parameters here represent
the global structure
of emerging magnetic flux
that eventually forms the AR,
and, since the HDF appears
only in the very initial phase
of the flux emergence,
the correlation with the HDF parameters
may not be so high.

\subsection{HDF and Elongated Granules}

In continuum,
the granulation pattern
in an emerging flux region
is known to appear
different from
that in quiet Sun.
In the early phase
of flux emergence,
the transient dark alignments,
or the darkened intergranular lanes,
appear
in the center
of the emerging region,
aligned roughly in parallel
with the axis connecting
the two main polarities
\citep{bra85,zwa85,str99}.
It is thought that
the dark lanes are created
by the horizontal magnetic fields
at the apex of the rising flux tube
from the convection zone
\citep{zwa85}.
Numerical simulations also
support this scenario
\citep[e.g.,][]{che07}

In event \#1
in Figure \ref{fig:clear1},
for example,
the granulation pattern
in the continuum image
looks mostly circular
at first at 07:15 UT
and also at 07:30 UT,
namely, after the HDF start
($t_{\rm HDF}=$ 07:24 UT).
However, at 08:00 UT,
the pattern in the central region
shows a slight elongation
to the direction
of the red and blue Doppler pair.
Although the flux emergence
is not detected yet
by the residual method
at this moment
($t_{\rm FE}=$ 08:11 UT),
the magnetogram shows
a faint positive (white) pattern,
which may be the horizontal magnetic fields
reflected because of
the projection
(this emerging AR is located
46.5$^{\circ}$ away
from the disk center).
However,
at 08:15 UT,
namely,
after the significant LoS flux
is detected at 08:11 UT,
the elongated pattern is not seen
in the continuum.
This transient elongation
reminds us of the concept
that the HDF is pushed up
by the horizontal magnetic fields
at the apex of the large-scale rising flux
transported from the deep convection zone,
which agrees well
with the ``two-step emergence'' scenario.




\section{Summary
  \label{sec:summary}}

In this paper,
we have shown
a statistical analysis
of newly emerging ARs.
In the numerical simulations
of the ``two-step emergence'' model
\citep[e.g.,][]{tor12a},
when the flux approaches
the solar surface,
unmagnetized plasma becomes trapped
between the rising flux and the photosphere
and eventually escapes horizontally
around the surface layer.
This HDF was previously
detected in a single emergence event
in \citet{tor12b}
and,
in the present study,
we extended the detection
in many more events,
aiming to investigate the characteristics
of the HDF.

Under the conditions of
(1) the solar minimum,
(2) $\theta\leq 60^{\circ}$,
and (3) the eastern hemisphere,
we picked up
23 flux emergence events
in 21 ARs,
total unsigned flux ranging
from $5.6\times 10^{20}$
to $2.3\times 10^{22}\ {\rm Mx}$.
Using the method
developed in \citet{tor12b},
we detected 6 clear HDFs.
In another 7 emergence events,
we found HDFs
by visual inspection.
In total,
the HDFs were observed
in 56.6\% of all events.
If we exclude the emergence events
closer to the disk center
($\theta\leq 30^{\circ}$),
which are supposed to have
less Doppler components,
the detection rate increases
up to more than 80\%.

In the 6 clear events,
the HDF duration
from the HDF appearance
to the flux emergence
was on average
61 minutes,
which is consistent with
103 minutes
observed in the previous detection
\citep{tor12b}
and 30--45 minutes
obtained in the numerical experiments
\citep{tor13a}.
According to the simulations,
this time gap is
comparable to
the waiting time
after the rising magnetic flux
slows down in the top convection zone
before it restarts emergence
into the upper atmosphere.
The maximum horizontal speed
of the HDF
in the present study
is on average
$3.1\ {\rm km\ s}^{-1}$,
which is also consistent with
$2.3\ {\rm km\ s}^{-1}$
observed in the previous detection
\citep{tor12b}
and several ${\rm km\ s}^{-1}$
in the simulations
\citep{tor13a}.

Assuming
a simple 2D model,
we estimated the rising speed
of the subsurface magnetic flux
that pushes up and drives the HDF.
The estimated rising speed
was $0.6$--$1.4\ {\rm km\ s}^{-1}$,
which is again consistent with
the simulations \citep{tor13a},
previous calculations
of the emergence
in the solar interior
\citep[e.g.,][]{fan09},
and other seismic studies
\citep[e.g.,][]{tor13b}.
By comparing with
the simulation results
in \citet{tor13a},
we also speculated that
the rising flux tubes
have a field strength of
less than $30\ {\rm kG}$
in the deeper convection zone
at around $-20\ {\rm Mm}$.

On the other hand,
it was found that
the correlations
between HDF parameters
(the duration and the flow speed)
and AR parameters
(the total flux,
its time derivative,
and the footpoint separation)
are not so high.
This may be because,
while the AR parameters represent
the global structure
of the rising magnetic fields,
the HDF parameters
are more focused on
the initial phase
of the flux emergence,
or the apex of the rising fields.
We also observed
the transient elongation
of granular cells
after the HDF was detected
in the emerging region.
Since the elongated structure
reflects the horizontal field
around the surface layer,
this observation is also
in good agreement with
the ``two-step emergence'' scenario.

In this analysis,
by comparing the temporal evolutions
of the magnetic and velocity distributions
with their reference quiet-Sun profiles,
we succeeded in detecting
the HDF signatures.
By visual inspection,
the HDF is easy to distinguish
from other convections
such as granulation and supergranulation
(as in Doppler images
in Figure \ref{fig:clear1}).
It is because,
although the typical size of the HDF
is about the same as that of supergranulation
($\sim 10$--$20\ {\rm Mm}$),
the horizontal speed is larger than
that of the supergranulation
(HDF $\sim 1.5\ {\rm km\ s}^{-1}$;
supergranulation $\sim$ a few $100\ {\rm m\ s}^{-1}$).
Moreover,
although the HDF velocity is comparable
to the granulation speed
($0.5$--$1.5\ {\rm km\ s}^{-1}$),
the size scale is by far different
(granulation $\sim 0.5$--$2\ {\rm Mm}$).
Given the above characteristics
of the HDF signatures,
together with our method that compares
with the reference quiet-Sun profiles,
one can see that
the strong flows
detected in this study
are not occurring by chance.
For obtaining
better statistical significance
of the HDF detection
and its quantification,
we can utilize
the full-disk Dopplergrams
in our future study
to expand the analyzed region
so as not to miss,
if any,
HDF events
without being associated
with magnetic field emergence
in the quiet-Sun regions.

In this paper,
we statistically analyzed the HDFs
in a larger ensemble
of emerging AR data.
We conclude here that
the HDF is
a rather common feature
in the earliest phase
of AR-scale flux emergence.
We also found that
the obtained HDF parameters
are highly consistent
with the numerical results
and the previous detection.
Moreover,
the HDF observation
provides us with a tool
to investigate the physical states
of the subsurface magnetic fields.



\acknowledgments

We thank the anonymous referee
for improving the manuscript.
S.T. is grateful to the {\it SDO} team
for distributing HMI data.
This work is based
on the PhD thesis
of S.T.
\citep[][Chapter 8]{tor14dis}.
S.T. was supported
by Grant-in-Aid for JSPS Fellows.






\appendix

\section{List of Target ARs
  \label{sec:catalog}}

Figure \ref{fig:catalog1}
is a list of target ARs
analyzed in this study.
Here,
we show
the HMI magnetogram,
the temporal evolutions
of the total unsigned flux $\Phi$
and its time derivative
(flux growth rate)
$d\Phi/dt$,
and the footpoint separation
$d_{\rm foot}$
between the two polarities.
The definition of each value
is shown
in Section \ref{sec:analysis:properties}.
In the magnetogram,
we also show the flux-weighted centers
of positive and negative polarities
with cross signs.
The flux and the centroids
are measured within the overlaid box.
Note that the physical values
of events \#22 and \#23
are not measured
because of the overlapping with each other.

\section{Clear HDF Events
  \label{sec:clear}}

Figure \ref{fig:clear1}
shows the temporal evolution
of 6 clear HDF events
(events \#1, \#2, \#3, \#9, \#15, and \#16)
observed by {\it SDO}/HMI.
Each column shows
the HMI continuum image,
Dopplergram, and magnetogram.
The overlaid square indicates
the area in which we plot the histogram
for the determination
of the HDF start $t_{\rm HDF}$
and the emergence start $t_{\rm FE}$.




\clearpage

\begin{figure}[tp]
  \begin{center}
    \includegraphics[width=140mm]{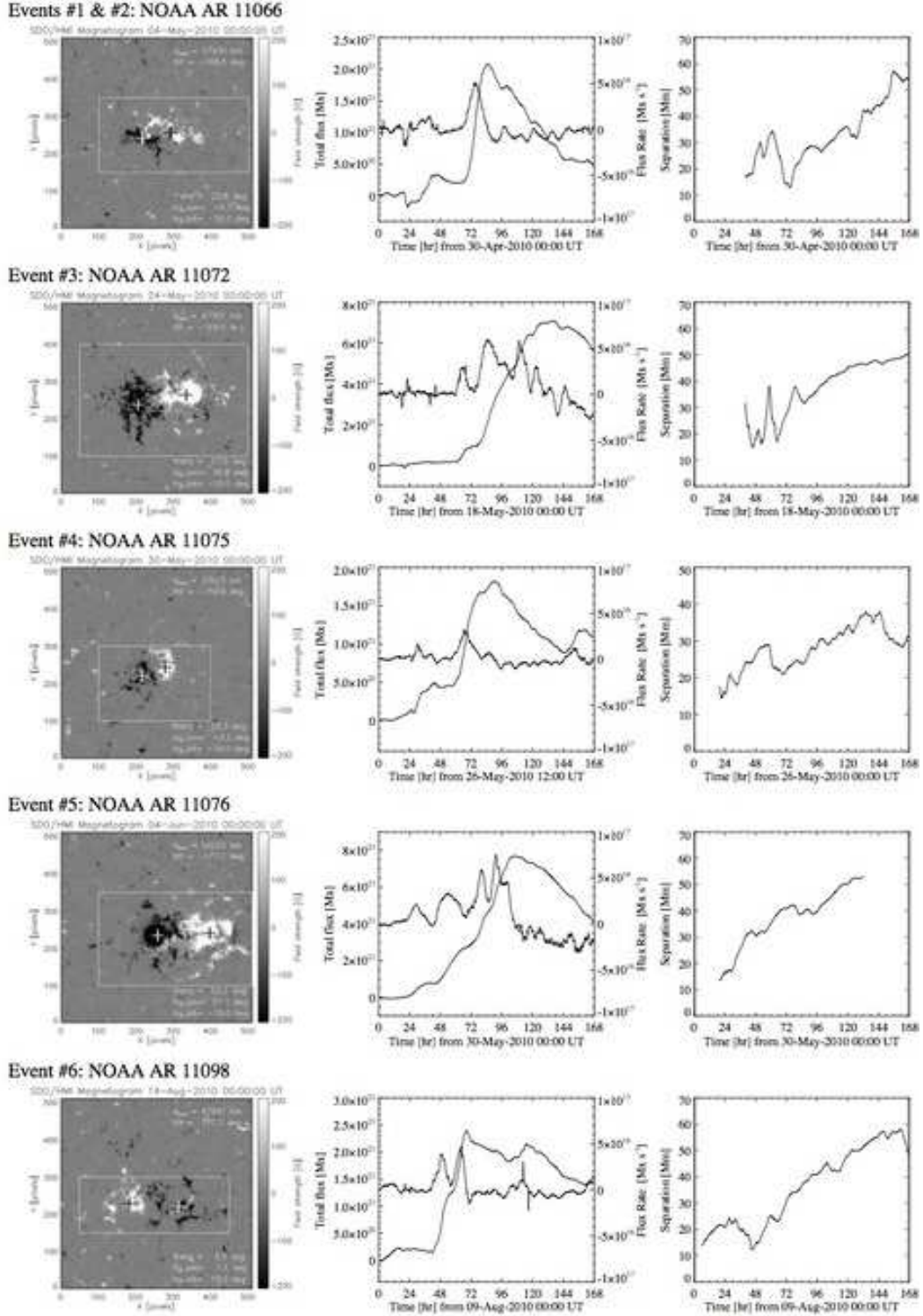}
  \end{center}
  \caption{
    ARs analyzed
    in this study.
    (Left) {\it SDO}/HMI magnetogram.
    Footpoint separation ($d_{\rm foot}$),
    tilt angle,
    heliocentric angle ($\theta$),
    heliographic latitude
    and longitude
    at the shown time
    are indicated.
    Black and white crosses
    denote the flux-weighted centers
    of the positive and negative polarities,
    respectively,
    which are measured
    within the box.
    (Middle) Temporal evolution
    of the total unsigned flux
    ($\Phi$: thick)
    and its time derivative
    ($d\Phi/dt$: thin).
    (Right) Footpoint separation
    ($d_{\rm foot}$).
  }
  \label{fig:catalog1}
\end{figure}

\addtocounter{figure}{-1}
\begin{figure}[tp]
  \begin{center}
    \includegraphics[width=140mm]{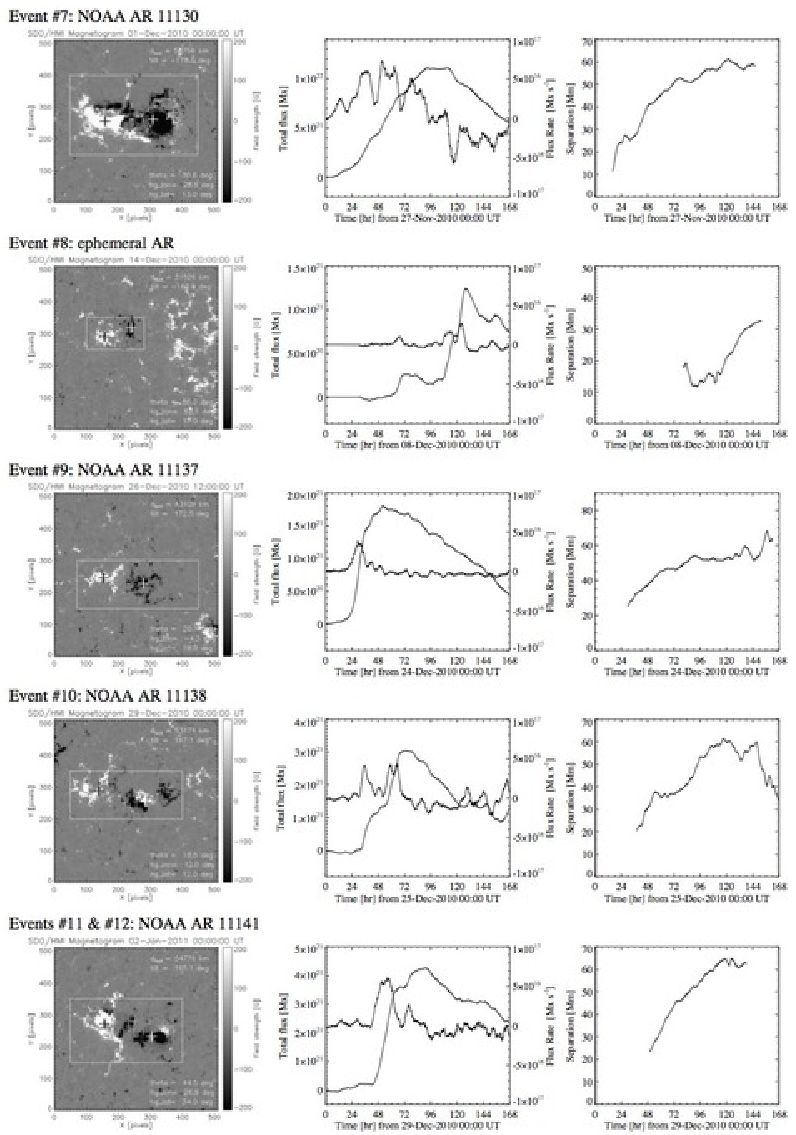}
  \end{center}
  \caption{{\it Continued.}}
  \label{fig:catalog2}
\end{figure}

\addtocounter{figure}{-1}
\begin{figure}[tp]
  \begin{center}
    \includegraphics[width=140mm]{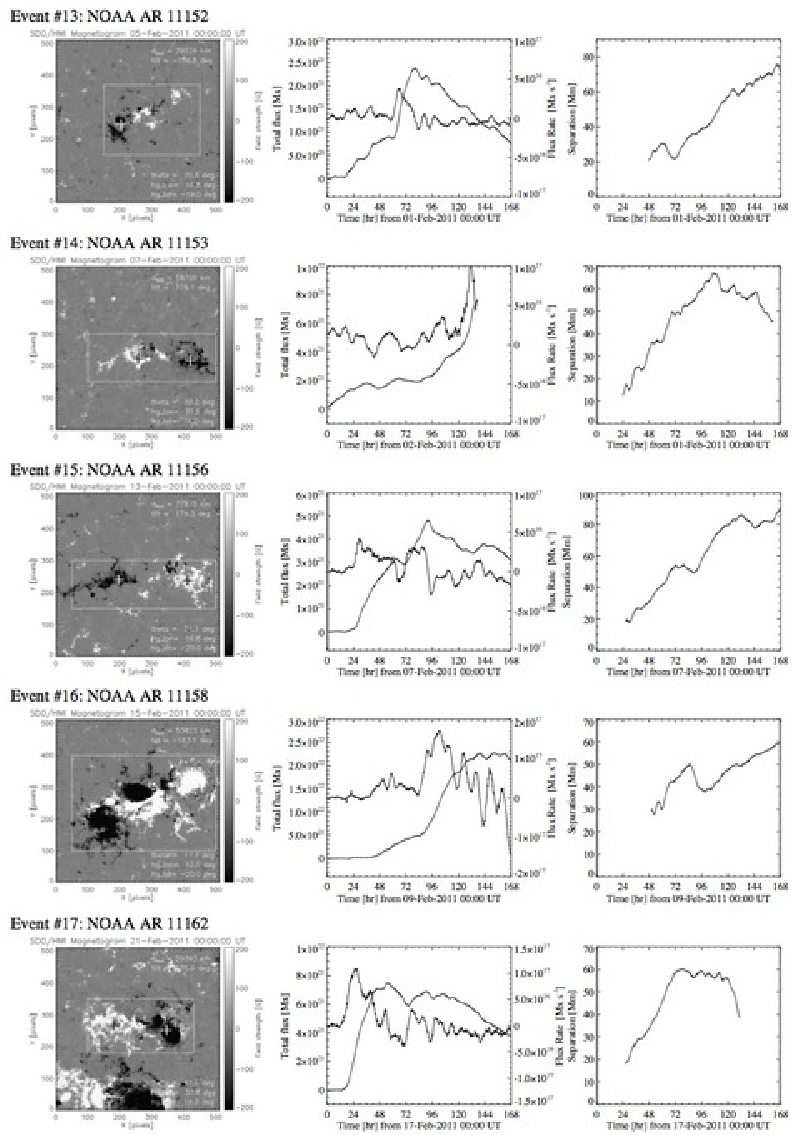}
  \end{center}
  \caption{{\it Continued.}}
  \label{fig:catalog3}
\end{figure}

\addtocounter{figure}{-1}
\begin{figure}[tp]
  \begin{center}
    \includegraphics[width=140mm]{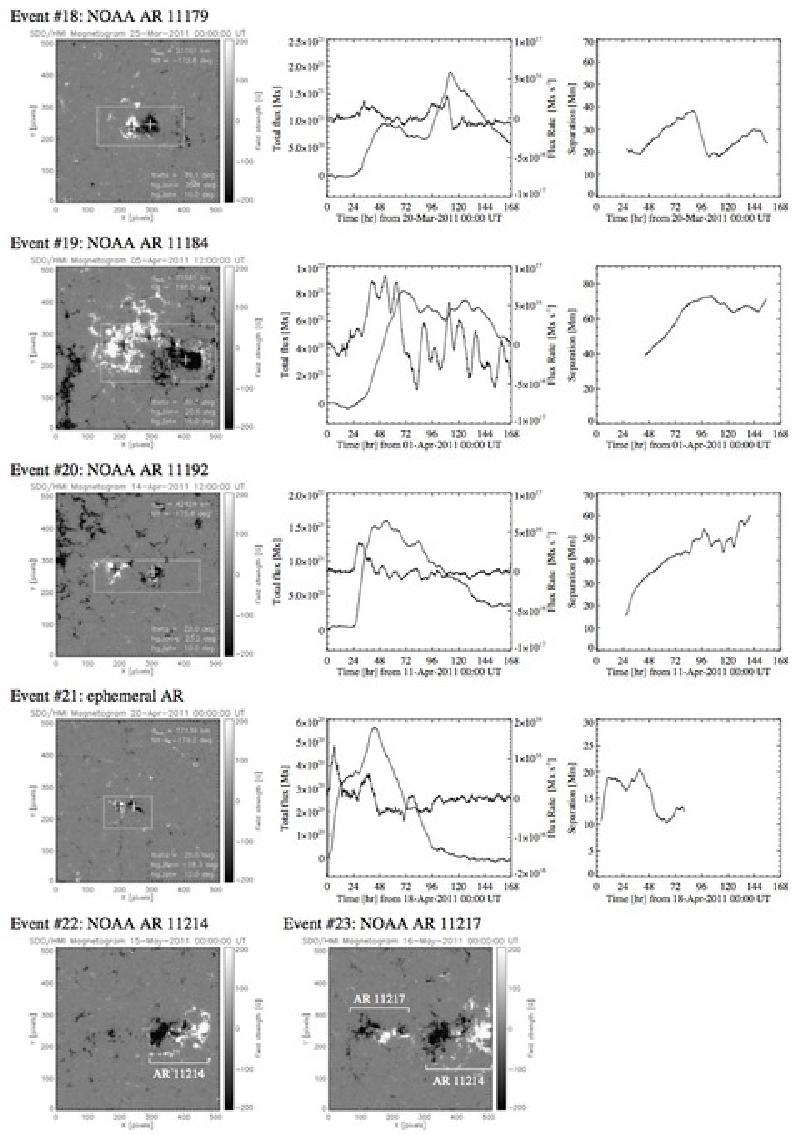}
  \end{center}
  \caption{{\it Continued.}}
  \label{fig:catalog4}
\end{figure}

\begin{figure}[tp]
  \begin{center}
    \includegraphics[width=140mm]{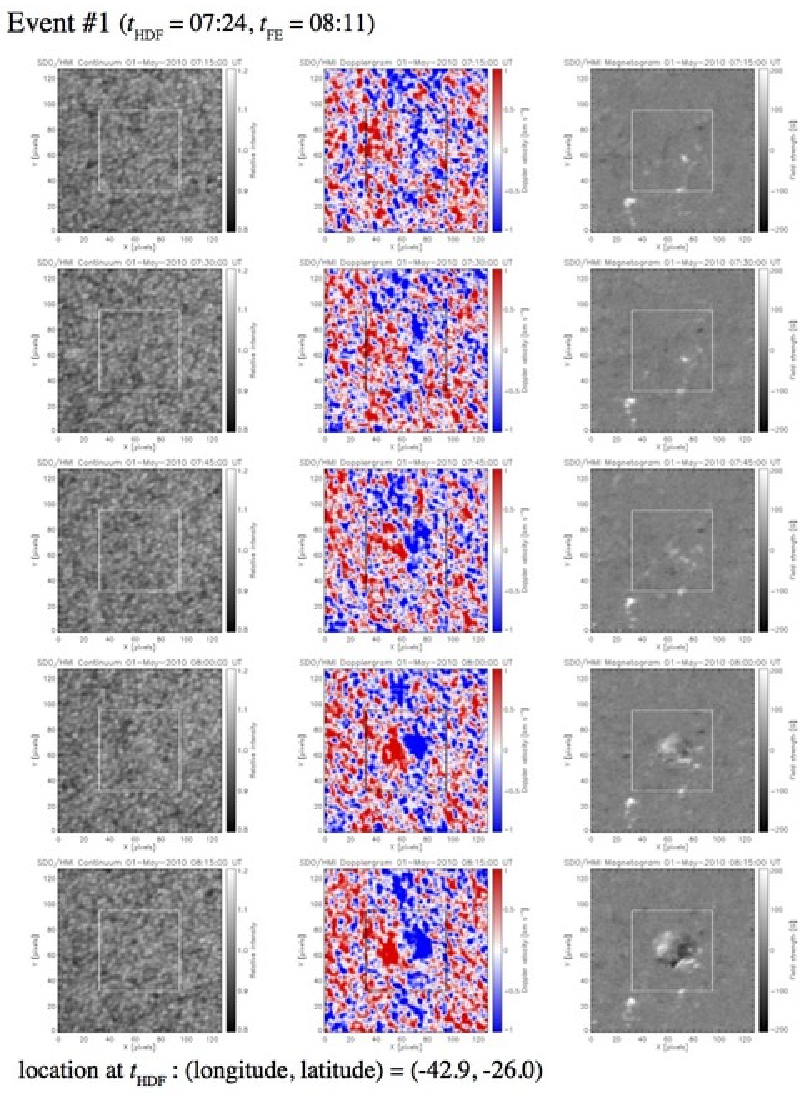}
  \end{center}
  \caption{
    Clear HDF events.
    Left, middle, and right columns
    show the evolutions
    of the continuum image,
    Dopplergram, and magnetogram,
    respectively.
    The square indicates
    the area in which
    we plot the histogram
    for the determination
    of the HDF start
    and the emergence start.
  }
  \label{fig:clear1}
\end{figure}

\addtocounter{figure}{-1}
\begin{figure}[tp]
  \begin{center}
    \includegraphics[width=140mm]{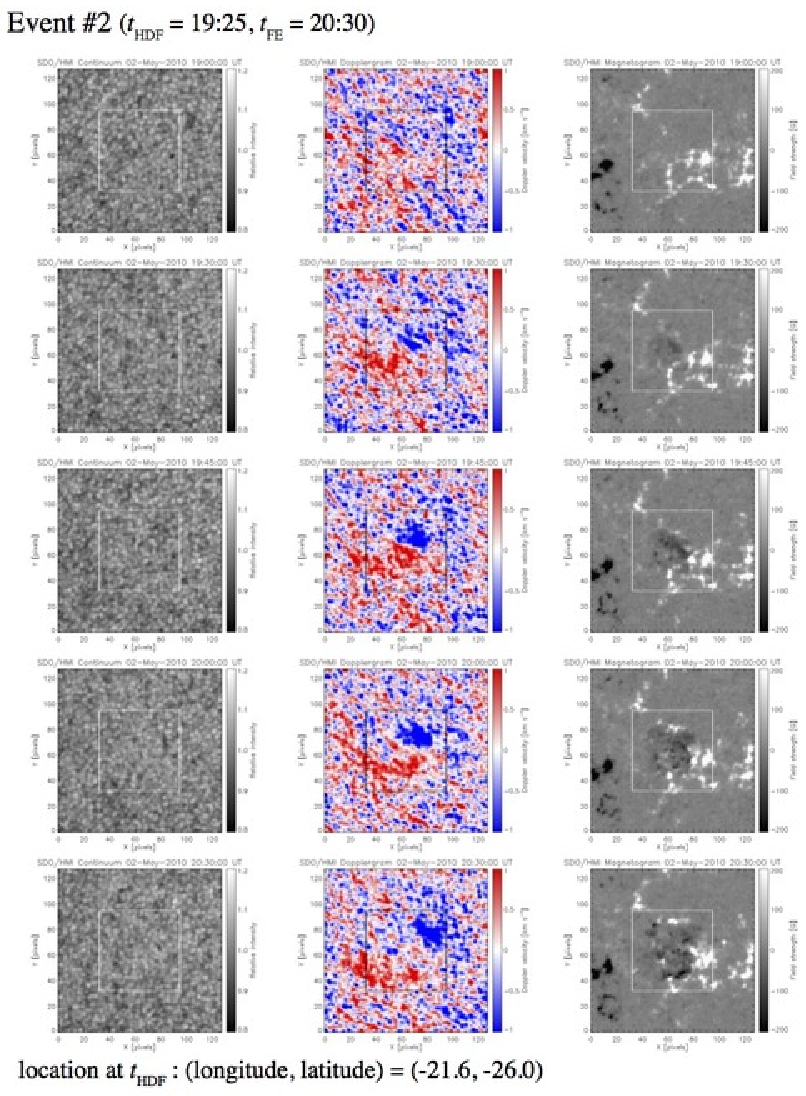}
  \end{center}
  \caption{{\it Continued.}}
  \label{fig:clear2}
\end{figure}

\addtocounter{figure}{-1}
\begin{figure}[tp]
  \begin{center}
    \includegraphics[width=140mm]{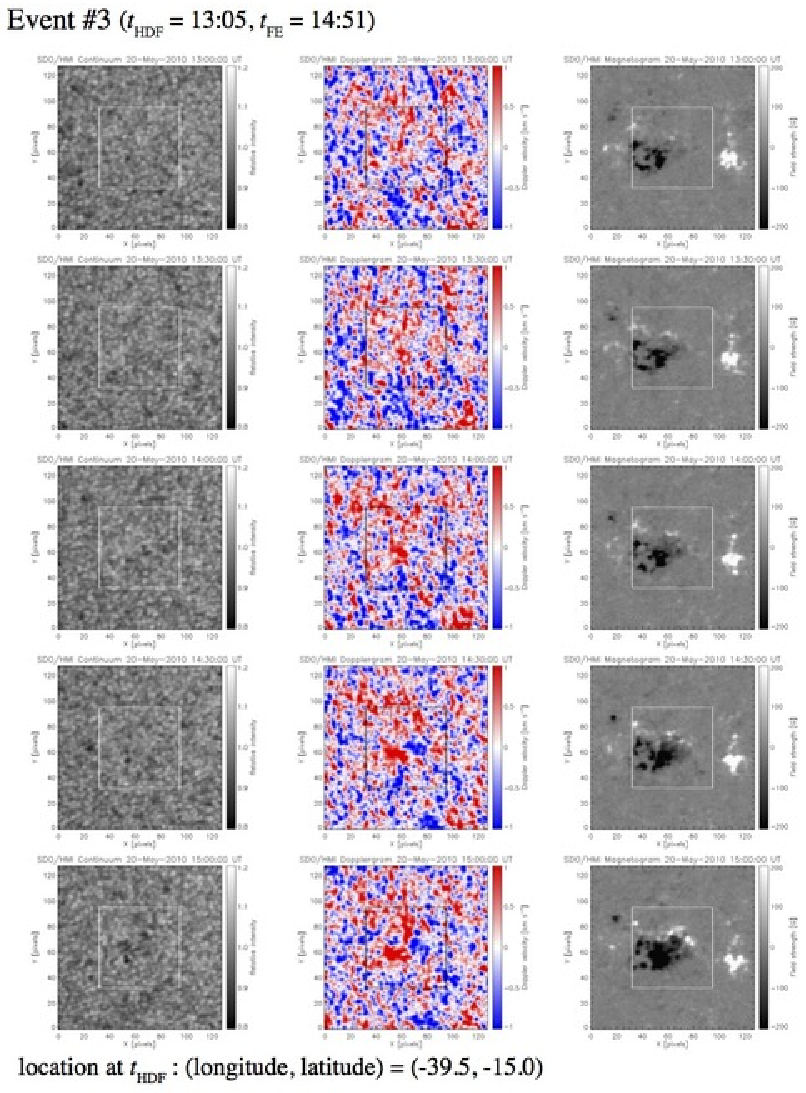}
  \end{center}
  \caption{{\it Continued.}}
  \label{fig:clear3}
\end{figure}

\addtocounter{figure}{-1}
\begin{figure}[tp]
  \begin{center}
    \includegraphics[width=140mm]{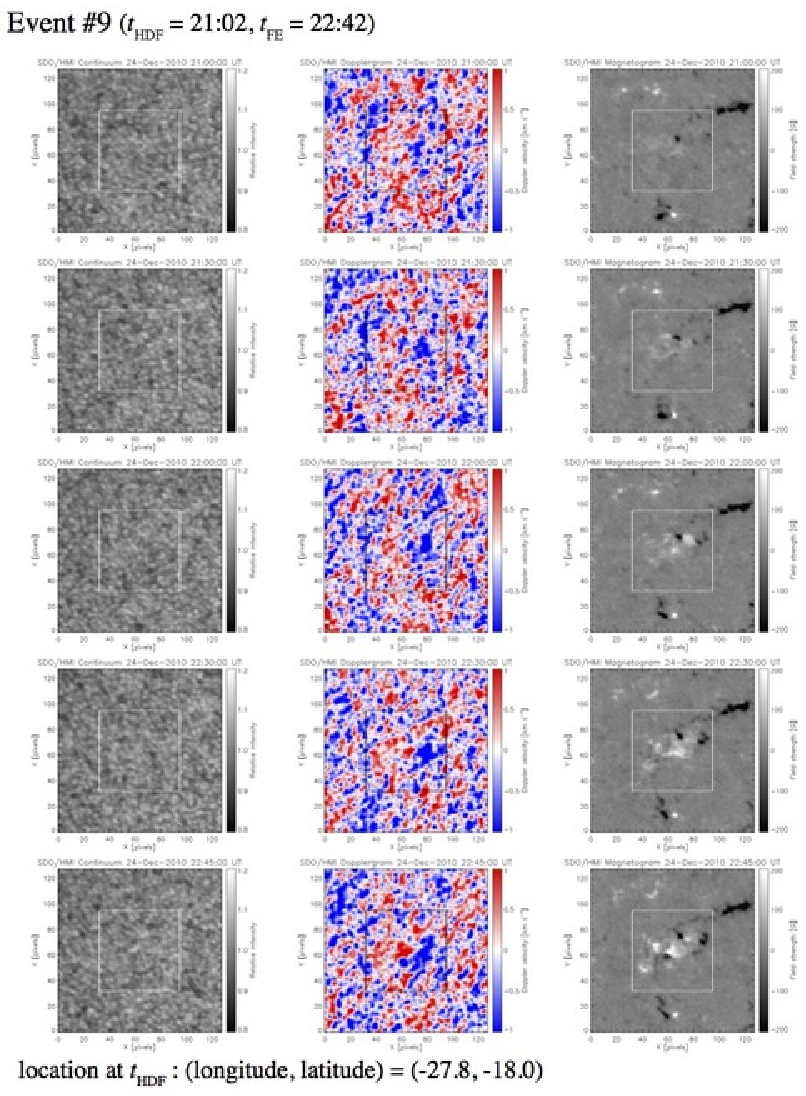}
  \end{center}
  \caption{{\it Continued.}}
  \label{fig:clear4}
\end{figure}

\addtocounter{figure}{-1}
\begin{figure}[tp]
  \begin{center}
    \includegraphics[width=140mm]{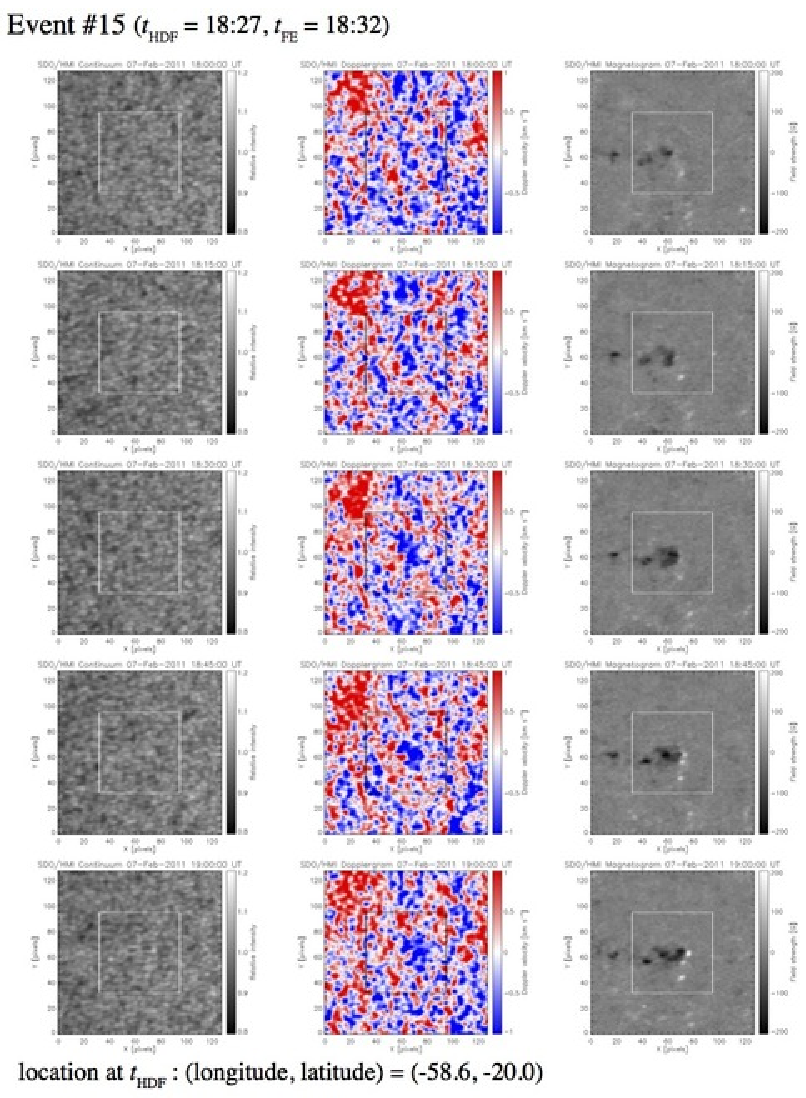}
  \end{center}
  \caption{{\it Continued.}}
  \label{fig:clear5}
\end{figure}

\addtocounter{figure}{-1}
\begin{figure}[tp]
  \begin{center}
    \includegraphics[width=140mm]{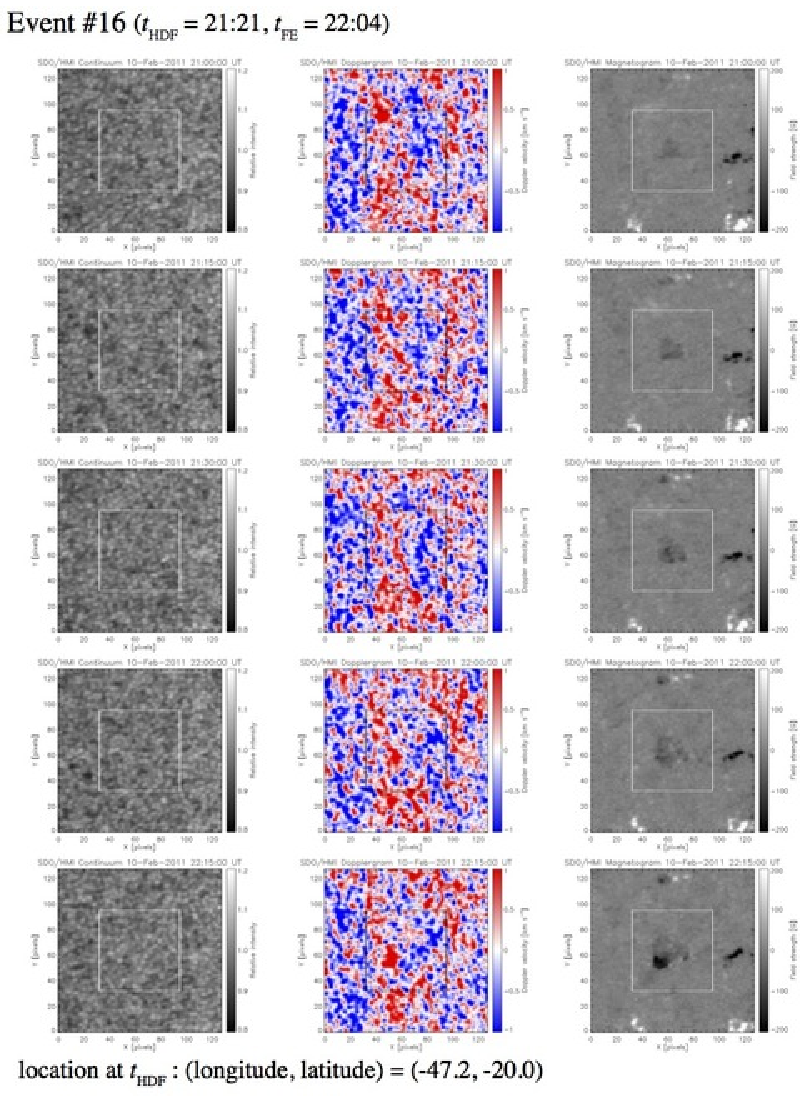}
  \end{center}
  \caption{{\it Continued.}}
  \label{fig:clear6}
\end{figure}

\clearpage
\begin{deluxetable}{lcllccc}
  \tabletypesize{\scriptsize}
  \tablecaption{Newly emerging ARs
    analyzed in this study.
    \label{tab:properties}}
  \tablewidth{0pt}
  \tablehead{
    \colhead{Event \#} & \colhead{NOAA AR \#} & \colhead{Year} &
    \colhead{Date} & \colhead{$\max{(\Phi)}$} & \colhead{$\max{(d\Phi/dt)}$} &
    \colhead{$\max{(d_{\rm foot})}$} \\
    \colhead{} & \colhead{} & \colhead{} & \colhead{} &
    \colhead{(Mx)} & \colhead{(Mx s$^{-1}$)} &
    \colhead{(Mm)}
  }
  \startdata
  1 & 11066 & 2010 & May 1 &
  $2.1\times 10^{21}$ & $5.1\times 10^{16}$ & $57.1$\\
  2\tablenotemark{a} & '' & & May 2 &
  '' & '' & '' \\
  3 & 11072 & & May 20 &
  $7.1\times 10^{21}$ & $6.0\times 10^{16}$ & $50.1$ \\
  4 & 11075 & & May 27 &
  $1.8\times 10^{21}$ & $3.3\times 10^{16}$ & $38.0$ \\
  5 & 11076 & & May 30 &
  $7.7\times 10^{21}$ & $7.5\times 10^{16}$ & $52.9$ \\
  6 & 11098 & & Aug 10 &
  $2.4\times 10^{21}$ & $4.7\times 10^{16}$ & $57.9$ \\
  7 & 11130 & & Nov 27 & $1.1\times 10^{22}$ & $7.3\times 10^{16}$ & $61.1$ \\
  8 & ephemeral & & Dec 10 &
  $1.2\times 10^{21}$ & $2.6\times 10^{16}$ & $32.6$ \\
  9 & 11137 & & Dec 24 &
  $1.8\times 10^{21}$ & $3.9\times 10^{16}$ & $68.3$ \\
  10 & 11138 & & Dec 26 &
  $3.1\times 10^{21}$ & $4.4\times 10^{16}$ & $61.4$ \\
  11 & 11141 & & Dec 29 &
  $4.3\times 10^{21}$ & $6.1\times 10^{16}$ & $64.8$ \\
  12\tablenotemark{b} & '' & & Dec 30 &
  '' & '' & '' \\
  13 & 11152 & 2011 & Feb 1 &
  $2.4\times 10^{21}$ & $3.8\times 10^{16}$ & $75.7$ \\
  14 & 11153 & & Feb 2 &
  $7.7\times 10^{21}$ & $1.1\times 10^{17}$ & $66.8$ \\
  15 & 11156 & & Feb 7 &
  $4.8\times 10^{21}$ & $4.2\times 10^{16}$ & $89.3$ \\
  16 & 11158 & & Feb 10 &
  $2.3\times 10^{22}$ & $1.7\times 10^{17}$ & $59.6$ \\
  17 & 11162 & & Feb 17 &
  $7.5\times 10^{21}$ & $1.1\times 10^{17}$ & $60.1$ \\
  18 & 11179 & & Mar 20 &
  $1.9\times 10^{21}$ & $3.1\times 10^{16}$ & $38.2$ \\
  19 & 11184 & & Apr ~2 &
  $8.2\times 10^{21}$ & $8.7\times 10^{16}$ & $72.8$ \\
  20 & 11192 & & Apr 12 &
  $1.6\times 10^{21}$ & $3.6\times 10^{16}$ & $59.8$ \\
  21 & ephemeral & & Apr 18 &
  $5.6\times 10^{20}$ & $1.3\times 10^{16}$ & $20.5$ \\
  22\tablenotemark{c} & 11214 & & May 13 & -- & -- & -- \\
  23\tablenotemark{c} & 11217 & & May 15 & -- & -- & -- \\
  \enddata
  \tablenotetext{a}{AR is the same as event \#1.}
  \tablenotetext{b}{AR is the same as event \#11.}
  \tablenotetext{c}{Physical values are not measured
    because of the overlapping of ARs 11214 and 11217
    (see Figure \ref{fig:catalog1}.)}
\end{deluxetable}

\begin{deluxetable}{lcllccclc}
  \tabletypesize{\scriptsize}
  \tablecaption{Results of the HDF detection.
    \label{tab:hdf}}
  \tablewidth{0pt}
  \tablehead{
    \colhead{Event \#} & \colhead{NOAA AR \#} & \colhead{Year} &
    \colhead{Date} & \colhead{$t_{\rm HDF}$\tablenotemark{a}} &
    \colhead{$t_{\rm FE}$\tablenotemark{b}} &
    \colhead{$\Delta t$\tablenotemark{c}} &
    \colhead{HDF\tablenotemark{d}} & \colhead{$\theta$}\\
    \colhead{} & \colhead{} & \colhead{} & \colhead{} &
    \colhead{} & \colhead{} & \colhead{(min)} &
    \colhead{} & \colhead{($^{\circ}$)}
  }
  \startdata
  1 & 11066 & 2010 & May ~1 & 07:24 & 08:11 & ~~47 & Y & 46.5 \\
  2 & '' & & May ~2 & 19:25 & 20:30 & ~~65 & Y & 30.1 \\
  3 & 11072 & & May 20 & 13:05 & 14:51 & ~106 & Y &40.6 \\
  4 & 11075 & & May 27 & -- & 05:59 & -- & N & 45.7 \\
  5 & 11076 & & May 30 & 17:08 & 16:31 & $-37$ & N & 35.7 \\
  6 & 11098 & & Aug 10 & (18:40) & 19:36 & ~~56 & Y$^{\ast}$ & 45.4 \\
  7 & 11130 & & Nov 27 & -- & 06:54 & -- & N & 28.0 \\
  8 & ephemeral & & Dec 10 & -- & 11:00 & -- & N & 17.2  \\
  9 & 11137 & & Dec 24 & 21:02 & 22:42 & ~100 & Y & 33.6 \\
  10 & 11138 & & Dec 26 & 10:21 & 07:49 & $-152$ & N & 29.7 \\
  11 & 11141 & & Dec 29 & -- & 13:44 & -- & N & 41.3 \\
  12 & '' & & Dec 30 & (19:15) & 20:08 & ~~53 & Y$^{\ast}$ & 36.7 \\
  13 & 11152 & 2011 & Feb ~1 & (16:41) & 16:58 & ~~52 & Y$^{\dagger}$ & 32.7 \\
  14 & 11153 & & Feb ~2 & (21:38) & 21:36 & ~~33 & Y$^{\dagger}$ & 35.7 \\
  15 & 11156 & & Feb ~7 & 18:27 & 18:32 & ~~~5 & Y & 58.2 \\
  16 & 11158 & & Feb 10 & 21:21 & 22:04 & ~~43 & Y & 47.4 \\
  17 & 11162 & & Feb 17 & (14:15) & 15:04 & ~~49 & Y$^{\ast}$ & 29.2 \\
  18 & 11179 & & Mar 20 & -- & 22:02 & -- & N & 29.6 \\
  19 & 11184 & & Apr ~2 & (00:40) & 02:21 & ~101 & Y$^{\ast}$ & 36.7 \\
  20 & 11192 & & Apr 12 & 02:48 & 01:19 & $-89$ & N & 19.9 \\
  21 & ephemeral & & Apr 18 & -- & 02:13 & -- & N & 49.0 \\
  22 & 11214 & & May 13 & (13:00) & 14:26 & ~~77 & Y$^{\dagger}$ & 43.6 \\
  23 & 11217 & & May 15 & -- & 07:29 & -- & N & 25.7\\
  \enddata
  \tablenotetext{a}{HDF start.
  Values determined by visual inspection
  are shown in parentheses.}
  \tablenotetext{b}{Emergence start.}
  \tablenotetext{c}{HDF duration:
    $\Delta t=t_{\rm FE}-t_{\rm HDF}$.}
  \tablenotetext{d}{Yes/No.
    Asterisk indicates the event
    of which $t_{\rm HDF}$ is not defined
    by the residual method
    but determined by visual inspection
    and $\Delta t>0$,
    while dagger indicates
    the event whose double-check result is negative
    but $t_{\rm HDF}$ is defined by eye
    and $\Delta t>0$.}
\end{deluxetable}

\begin{deluxetable}{lcccccc}
  \tablecaption{Clear HDF events.\label{tab:clear}}
  \tablewidth{0pt}
  \tablehead{
    \colhead{Event \#} & \colhead{$\Delta t$} & \colhead{$\theta$} &
    \colhead{$\max(|V_{\rm D}|)$} & \colhead{$\max(|V_{\rm h}|)$} &
    \colhead{$\max(d\Phi/dt)$} & \colhead{$V_{z}$\tablenotemark{a}}\\
    \colhead{} & \colhead{(min)} & \colhead{($^{\circ}$)} &
    \colhead{(km s$^{-1}$)} & \colhead{(km s$^{-1}$)} &
    \colhead{(Mx s$^{-1}$)} & \colhead{(km s$^{-1}$)}
  }
  \startdata
  1 & 47 & 46.5 & 2.8 & 3.9 & $1.4\times 10^{16}$ & 1.3 \\
  2 & 65 & 30.1 & 2.2 & 4.4 & $2.1\times 10^{16}$ & 1.4 \\
  3 & 106 & 40.6 & 2.4 & 3.7 & $1.1\times 10^{16}$ & 1.2 \\
  9 & 100 & 33.6 & 1.7 & 3.1 & $9.4\times 10^{15}$ & 1.0 \\
  15 & 5 & 58.2 & 1.5 & 1.8 & $1.2\times 10^{16}$ & 0.6 \\
  16 & 43 & 47.4 & 1.5 & 2.0 & $3.0\times 10^{15}$ & 0.7 \\
  \enddata
  \tablenotetext{a}{This quantity is calculated
    using Equation (\ref{eq:masscons}),
    rather than derived directly from the data.
    See text for details.}
\end{deluxetable}


\end{document}